\documentclass[amsmath,jcp,superscriptaddress,twocolumn]{revtex4-1}
\usepackage{float}
\usepackage{color}
\usepackage{amsmath}
\usepackage{amsfonts}
\usepackage{amssymb}
\usepackage{MnSymbol}
\usepackage{graphicx}
\usepackage[caption=false]{subfig}
\usepackage{tikz}
\usetikzlibrary{arrows}

\usepackage{tikz-feynman}
\tikzfeynmanset{compat=1.0.0}

\begin{document}

\title{Two-dimensional spectroscopy of open quantum systems}

\author{Haoran Sun}
\affiliation{Department of Chemistry \& Biochemistry, University of California San Diego, La Jolla, CA 92093, USA}
\author{Upendra Harbola}
\affiliation{Department of Inorganic and Physical Chemistry, Indian Institute of Science, Bangalore 560012, India}
\author{Shaul Mukamel}
\affiliation{Department of Chemistry, University of California Irvine, Irvine, CA 92697, USA}
\author{Michael Galperin}
\email{mgalperin@tauex.tau.ac.il}
\affiliation{School of Chemistry, Tel Aviv University, Tel Aviv 69978, Israel}


\begin{abstract}
Two-dimensional spectroscopy is discussed for open quantum systems
with multiple simultaneously measurable fluxes. In particular, we discuss
a junction where optical measurements of photon flux are complemented
with simultaneous transport measurements of electron currents.
Theory of two-dimensional spectroscopy in both fluxes is developed
employing non-self-consistent nonequilibrium Green's function
formulation. Theoretical derivations are illustrated with numerical 
simulations within generic junction model. 
\end{abstract}

\maketitle


\section{Introduction}
Interaction of light with matter is one of central research areas in chemistry.
Traditionally, spectroscopy is a tool to investigate responses of a system
to external perturbation by radiation field. Originally, studies were focused
on molecules in gas phase. Later, with development of experimental techniques,
optical spectroscopy studies were extended to open molecular systems
(molecules at surface, molecules at interface, molecules in cavities)~\cite{benz_nanooptics_2015,chikkaraddy_single-molecule_2016,chikkaraddy_single-molecule_2023}.
Contrary to the gas phase experiments, photon flux is not the only 
response measurable in such systems. Electron transfer and transport 
is one more measurable characteristic in open molecular systems.
Similar to optical spectroscopy with its focus on photon flux,
current spectroscopy uses bias induced electron flux to study 
properties of molecules. For example, such is resonant~\cite{zhitenev_conductance_2002,leroy_electrical_2004} 
and off-resonant~\cite{bayman_shifts_1981,hahn_electronic_2000,wang_inelastic_2004,smit_measurement_2002,agrait_onset_2002} inelastic electron tunneling spectroscopy.
Research focused on combination of light (photon) and matter (electron) responses 
in single molecule junctions is called optoelectronics~\cite{galperin_molecular_2012,galperin_photonics_2017}.

Molecular optoelectronics encompasses several research directions originally
focused on one type of response due to the other type of perturbation.
For example, such is bias induced light~\cite{imada_real-space_2016,imada_single-molecule_2017,miwa_many-body_2019} including 
light emission above threshold when emitting photon energy energy 
is higher that energy available due to bias~\cite{schull_electron-plasmon_2009,schneider_optical_2010,zhu_hot-carrier_2020}.
The opposite example is light induced current (photocurrent)~\cite{noy_response_2010,vadai_plasmon-induced_2013,vadai_plasmon-induced_2016,miwa_hubbard_2019,imai-imada_orbital-resolved_2022}
including control of the current by light~\cite{arielly_accurate_2011,wang_quantum_2019,muniz_quantum_2019}.
Further development of laser techniques at nanoscale yield possibility of
simultaneous measurement of photon and electron fluxes in single molecule
junctions~\cite{ioffe_detection_2008,ward_simultaneous_2008,ward_vibrational_2011,natelson_nanogap_2013,li_voltage_2014,jaculbia_single-molecule_2020}.

\begin{figure}[b]
\centering\includegraphics[width=\linewidth]{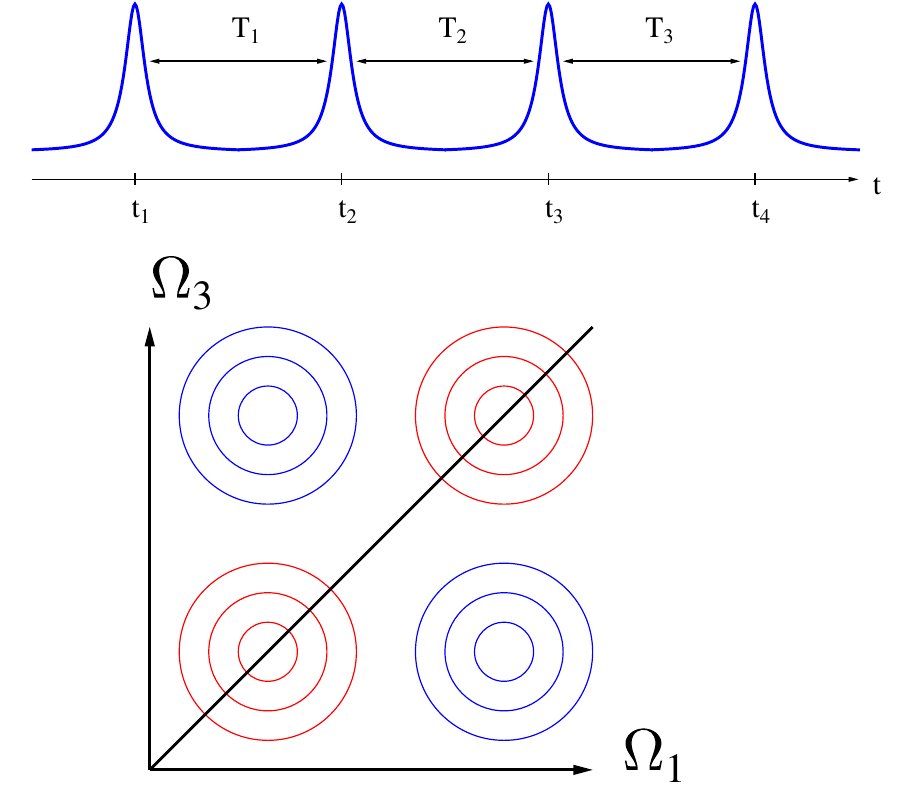}
\caption{\label{fig1}
 Sketch of two-dimensional spectroscopy signal.
}
\end{figure}

Multidimensional optical spectroscopy is used as a tool to study electronic structure 
and dynamics of complex molecular systems in condensed phase. 
Measurements extending the technique to electron flux (two-dimensional photocurrent
spectroscopy) were reported in the literature a decade ago~\cite{nardin_multidimensional_2013,karki_coherent_2014}.
Multidimensional spectroscopy with simultaneous measurement of
electron and photon fluxes is the natural next step.
Standard experimental setup 
uses sequence of four laser pulses while measuring
response of the system to the perturbation as function of time-delays
$T_1$, $T_2$, and $T_3$ between pulses. 
The response is recorded as 2d diagram of integrated flux
vs. $\Omega_1$ and $\Omega_3$ (Fourier transforms of sequences of 
measurements in delays $T_1$ and $T_3$) which gives information on
quantum coherences in the system. Sequence of diagrams for different
$T_2$ yields information on dynamics of the coherences (see Fig.~\ref{fig1}).

Here, we present theoretical description of two-dimensional optical
spectroscopy in open quantum systems when both photon and electron
fluxes are measured simultaneously. Our recent formulation of nonlinear 
spectroscopy in open quantum systems~\cite{mukamel_nonlinear_2024}
is used as a starting point. Quantum description of radiation field is 
reduced to classical treatment of laser pulses, connection to
Feynman diagrams of quantum treatment is established.
We derive explicit expressions for the multidimensional signals
due to photon and electron fluxes following standard formulation
of nonlinear optical spectroscopy, where the signal is modeled 
as fourth order response function. Finally, we illustrate limitations
of the approach and discuss possible alternatives.

The paper is organized as follows. After introducing a model of 
an open quantum system and discussing treatment of its responses 
within the nonequilibirum Green's function (NEGF) method,
we derive expressions for two-dimensional signals obtained
by methods of nonlinear spectroscopy for photon and electron fluxes.
Simple generic models are employed to present numerical
results  illustrating our findings. Our presentation is finalized 
with conclusions and future research directions.


\section{Model}\label{model}
We consider molecular system, $M$, coupled to two contacts, $L$ and $R$,
each at its own equilibrium, and driven by external classical field $E(t)$.
Hamiltonian of the model is 
\begin{equation}
\label{H}
\hat H(t) = \hat H_M(t) + \sum_{K=L,R}\left(\hat H_K+\hat V_{MK}\right)
\end{equation}
where $\hat H_M(t)$ and $\hat H_K$ describe, respectively
driven molecular system and contact $K$. $\hat V_{MK}$ couples
between the subsystems. Explicit expressions for the Hamiltonians
are
\begin{align}
\label{HM}
\hat H_M(t) &= \hat H_M^{(0)} -\hat P\,E(t)
\\
\label{HK}
\hat H_K &= \sum_{k\in K}\varepsilon_k\hat c_k^\dagger\hat c_k
\\
\label{VKM}
\hat V_{KM} &= \sum_{k\in K}\sum_{m\in M}\left(V_{km}\hat c_k^\dagger\hat d_m +\mbox{H.c.}\right)
\end{align} 
Here, $\hat d_m^\dagger$ ($\hat d_m$) and $\hat c_k^\dagger$ ($\hat c_k$)
creates (annihilates) electron in molecular orbital $m$ and state $k$ of contacts,
respectively.
\begin{equation}
\hat P\equiv \sum_{m,n \in M}\mu_{mn}\hat d_m^\dagger\hat d_n
\end{equation}
is the molecular dipole operator.
$\hat H_M^{(0)}$ describes molecular system in the absence of
driving.

We analyze electron $I^K_e(t)$ ($K=L,R$) and photon $I_p(t)$ fluxes. 
Their explicit expressions are~\cite{jauho_time-dependent_1994,mukamel_nonlinear_2024}
\begin{align}
\label{IK}
I_e^K(t) &= 2\,\mbox{Re}\sum_{m,m'\in M}\int_{-\infty}^t dt'\,\bigg(
\Sigma^K_{mm'}(t,t')\, G^{>}_{m'm}(t',t)
-\Sigma^K_{mm'}(t,t')\, G^{<}_{m'm}(t',t) \bigg)
\\
\label{Ip}
I_p(t) &= -2\,\mbox{Re}\sum_\alpha \lvert E_\alpha\rvert^2\int_{-\infty}^t dt'\,\bigg(
F_\alpha^{(0)\, <}(t,t')\, G^{>}_{PP}(t',t) 
- F_\alpha^{(0)\, >}(t,t')\, G_{PP}^{<}(t',t)
\bigg)
\end{align}
Here, $G$ and $G_{PP}$ are the single and two-particle Green's functions
\begin{align}
\label{G}
G_{m_1m_2}(\tau_1,\tau_2) &\equiv -i\langle T_c\,\hat d_{m_1}(\tau_1)\,\hat d_{m_2}^\dagger(\tau_2)\rangle
\\
\label{GPP}
G_{PP}(\tau_1,\tau_2) &= -i\langle T_c\, \hat P(\tau_1)\,\hat P^\dagger(\tau_2)\rangle
\end{align}
and $\Sigma^K$ is the electron self-energy due to coupling to contact $K$
\begin{equation}
\label{SK}
\Sigma^K_{m_1m_2}(\tau_1,\tau_2) \equiv \sum_{k\in K} V_{m_1k}\,
g_k(\tau_1,\tau_2)\, V_{km_2}
\end{equation}
where
\begin{equation}
g_k(\tau_1,\tau_2) \equiv -i\langle T_c\,\hat c_k(\tau_1)\,\hat c_k^\dagger(\tau_2)\rangle_0
\end{equation}
is the Green's function of free electron in state $k$ of the contacts.

Note that, similar to standard nonlinear optical spectroscopy, 
derivation of  photon flux is based on quantum description of radiation field. 
Correspondingly, $F^{(0)}_\alpha$ in (\ref{Ip}) is
Green's function of free photons in mode $\alpha$ of the field,
\begin{equation}
 F^{(0)}_\alpha(\tau_1,\tau_2)\equiv -i\langle T_c\,\hat a_\alpha(\tau_1)\hat a_\alpha^\dagger(\tau_2)\rangle,
\end{equation} 
and $\hat a_\alpha^\dagger$ ($\hat a_\alpha$) creates (destroys) photon in
the mode $\alpha$. 

In our model, Eq.(\ref{H}), radiation field is treated classically (as external
driving force). Correspondence between quantum and classical descriptions
can be obtained by the following substitution
\begin{equation}
\label{QM2CM}
\lvert E_\alpha\rvert^2 F_\alpha(\tau_1,\tau_2)
\rightarrow -i\, \mathcal{E}_\alpha(t_1)\,\mathcal{E}_\alpha^{*}(t_2)
\end{equation} 
where $t_{1,2}$ are physical times corresponding to contour variables 
$\tau_{1,2}$ and 
\begin{equation}
\mathcal{E}_\alpha(t)\equiv E_\alpha e^{-i\omega_\alpha t+i\phi_\alpha} 
\end{equation}
is value of component of classical field mode $\alpha$ at time $t$.
We note that not all Feynman diagrams of quantum consideration 
contribute to classical radiation field treatment. For example, diagrams
corresponding to virtual quantum processes should not be taken into account.
Indeed, classical field does not support virtual processes.
This is reflected in the fact that, e.g. fourth order diagram in Fig.~6c of 
Ref.~\cite{mukamel_nonlinear_2024} splits into two independent
processes of second order when transferring to classical description.


\section{Two-dimensional spectroscopy}\label{cur2d}
Two-dimensional spectroscopy is based on response of system
to sequence of four monochromatic pulses at ordered set of times 
$t_1<t_2<t_3<t_4$ (see Fig.~\ref{fig1}). 
Traditional spectroscopy treats photon flux as fourth order process in
light-matter interaction (third order response function)~\cite{mukamel_principles_1995}.

Analog of such treatment in open systems starts by writing fourth order
expressions for electron and photon fluxes, Eqs.~(\ref{IK}) and (\ref{Ip}), 
within the  the NEGF bare diagrammatic expansion as introduced in 
Ref.~\cite{mukamel_nonlinear_2024}.
They are obtained by substituting fourth order contribution to $G$
into (\ref{IK}) and second order contribution of $G_{PP}$ into (\ref{Ip})
(see Eqs.~(12), (28), (A1), (A2), and (B1) of Ref.~\cite{mukamel_nonlinear_2024}).
Explicit expressions for the fluxes when radiation field is treated classically 
are given in Appendix~\ref{app_flux}.

\begin{figure}[htbp]
\centering\includegraphics[width=\linewidth]{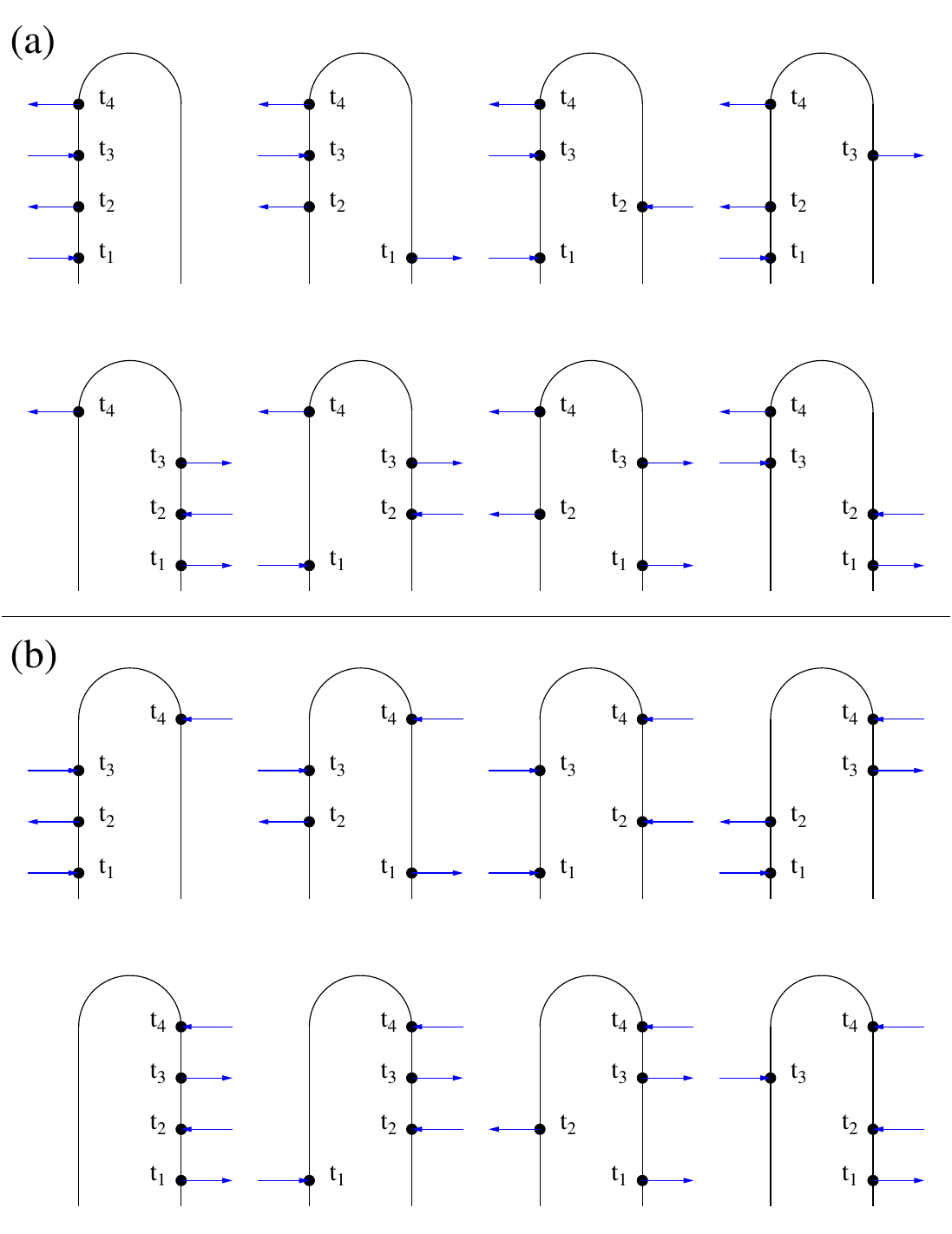}
\caption{\label{fig2}
 Projections (double-sided Feynman diagrams) contributing 
 to the two-dimensional signal.
}
\end{figure}

System response in the two-dimensional spectroscopy is characterized by 
the Fourier transform in delay times $T_1$ and $T_3$ (for fixed $T_2$)
of integrated flux 
\begin{align}
\label{SOm}
& S(\Omega_1,T_2,\Omega_3) =
 \int_0^\infty dT_1\int_0^\infty dT_3\,
S(T_1,T_2,T_3)\, e^{i\Omega_1T_1+i\Omega_3 T_3}
\end{align}  
Here, $S(T_1,T_2,T_3)$ is total charge passed at interface $K$ 
or total number of photons
\begin{align}
\label{SK}
S_e^K(T_1,T_2,T_3) &\equiv \int_{-\infty}^{+\infty} dt\, I_e^K(t)
\\
\label{Sp}
S_p(T_1,T_2,T_3) &\equiv \int_{-\infty}^{+\infty} dt\, I_p(t)
\end{align}
Evaluation of explicit expressions is based on several observations
and assumptions
\begin{enumerate}
\item One has to establish connection between
$t_a, t_b, t_c, t_d$ (physical times corresponding to contour variables 
$\tau_a,\tau_b,\tau_c,\tau_d$) in expressions for the fluxes, 
Eqs.~(\ref{IK4}) and (\ref{Ip4}), and $t_1,t_2,t_3,t_4$ of the pulses
(see Fig.~\ref{fig1}). For photon flux $I_p(t)$, it is natural to identify $t_4$
with the latest time of the signal $t$, then $t_2$ corresponds to $t_b$
while pair $t_1, t_3$ can be identified with $t_a$ and $t_c$ in any ordering.
For electron flux $I_e^K(t)$, pair $t_4,t_2$ is identified with
$t_b$ and $t_b$ in any order, while pair $t_3,t_1$ corresponds to 
$t_c$ and $t_a$ also in any ordering. 
\item Contour integrations in Eqs.~(\ref{IK4}) an (\ref{Ip4})
imply all possible placements of variables $\tau_{a,b,c,d}$ on the 
Keldysh contour. However, condition $t_4>t_3>t_2>t_1$
restricts placement possibilities to projections 
(double-sided Feynman diagrams) presented in Fig.~\ref{fig2}.
Note that for photon flux only projections in Fig.~\ref{fig2}a (or only in Fig.~\ref{fig2}b)
have to be taken into  account. 
This follows from the choice $t=t_4$ and cyclic property of trace.
Note also, that for electron flux there is also integration in $\tau'$.
Thus, for each line in expression (\ref{IK4}) one has to consider 
$3\times 2^4=48$ projections, each line in Eq.(\ref{Ip4}) yields $2^3=8$
projections.
\item To simplify the expressions spectroscopic analysis assumes 
laser pulses to be local in time (width of the pulse is zero).
This removes corresponding integrals in the final results.
\end{enumerate}

\begin{widetext}
Using  (\ref{IK4})-(\ref{Ip4}) 
in (\ref{SOm})-(\ref{Sp}) 
and taking into account observations mentioned above
yields 
\begin{align}
\label{SKOm}
& S_e^K(\Omega_1,T_2,\Omega_3) =
-2\,\mbox{Im}\int\frac{d\epsilon_1}{2\pi}\int\frac{d\epsilon_2}{2\pi}\, 
E_4^{*}\, E_3\, E_2^{*}\, E_1\,
e^{iT_2(\omega_1-\omega_2+\epsilon_1-\epsilon_2)+i\Delta\phi}\,
\mbox{Tr}\bigg\{
\\ & 
\,\,\,\,\,\big[G^{(0)<}(\epsilon_2)\,\mu^\dagger\, G^{(0)>}(\epsilon_2-\Omega_1-\omega_1)
- (>\,\leftrightarrow\, <)
\big]
\mu\, A^{(0)}(\epsilon_1)\,\mu^\dagger\, A^{(0)}(\epsilon_2-\Omega_3-\omega_4)\,
\mu
\nonumber \\ & \qquad\qquad\qquad\times
\left[A^{(0)}(\epsilon_2)\,\Sigma_K^{a}(\epsilon_2)
+G^{(0)r}(\epsilon_2)\,i\left(\Sigma_K^{>}(\epsilon_2)-\Sigma_K^{<}(\epsilon_2)\right)
\right]
\nonumber \\ & - 
A^{(0)}(\epsilon_2)\,\mu\left[
G^{(0)<}(\epsilon_1+\Omega_1+\omega_1)\,\mu^\dagger\, G^{(0)>}(\epsilon_1)
- (>\,\leftrightarrow\, <)
\right]\mu^\dagger\, 
A^{(0)}(\epsilon_2-\Omega_3-\omega_4)\,\mu
\nonumber \\ & \qquad\qquad\qquad\times
\left[A^{(0)}(\epsilon_2)\,\Sigma_K^{a}(\epsilon_2)+
G^{(0)r}(\epsilon_2)\, i\left(\Sigma_K^{>}(\epsilon_2)-\Sigma_K^{<}(\epsilon_2)\right)
\right]
\nonumber \\ & - 
A^{(0)}(\epsilon_1+\Omega_3+\omega_4)\,\mu^\dagger\left[
G^{(0)<}(\epsilon_2)\,\mu^\dagger\, G^{(0)>}(\epsilon_2-\Omega_1-\omega_1)
- (>\,\leftrightarrow\, <)
\right]\mu\, A^{(0)}(\epsilon_1)\,\mu
\nonumber \\ & \qquad\qquad\qquad\times
\big[
A^{(0)}(\epsilon_1+\Omega_3+\omega_4)\,\Sigma_K^{a}(\epsilon_1+\Omega_3+\omega_4)
+G^{(0)r}(\epsilon_1+\Omega_3+\omega_4)\, 
\nonumber \\ & \qquad\qquad\qquad\qquad\times
i\left(
\Sigma_K^{>}(\epsilon_1+\Omega_3+\omega_4) -
\Sigma_K^{<}(\epsilon_1+\Omega_3+\omega_4) \right)
\big]
\nonumber \\ & + 
A^{(0)}(\epsilon_1-\Omega_3-\omega_4)\,\mu^\dagger\, A^{(0)}(\epsilon_2)\,\mu
\left[ G^{(0)<}(\epsilon_1-\Omega_1-\omega_1)\,\mu^\dagger\, G^{(0)>}(\epsilon_1)
- (>\,\leftrightarrow\, <)\right]\mu
\nonumber \\ & \qquad\qquad\qquad\times
\big[
A^{(0)}(\epsilon_1-\Omega_3-\omega_4)\,\Sigma_K^{a}(\epsilon_1-\Omega_3-\omega_4)
+G^{(0)r}(\epsilon_1-\Omega_3-\omega_4)\, 
\nonumber \\ & \qquad\qquad\qquad\qquad\times
i\left(
\Sigma_K^{>}(\epsilon_1-\Omega_3-\omega_4) -
\Sigma_K^{<}(\epsilon_1-\Omega_3-\omega_4)\right)
\big]
\nonumber \\ & - 
A^{(0)}(\epsilon_2-\Omega_3-\omega_4)\,\mu\left[
G^{(0)<}(\epsilon_2)\,\mu^\dagger\, G^{(0)>}(\epsilon_2-\Omega_1-\omega_1)
- (>\,\leftrightarrow\, <)
\right]\mu\, A^{(0)}(\epsilon_1)\,\mu^\dagger
\nonumber \\ & \qquad\qquad\qquad\times
\big[
A^{(0)}(\epsilon_2-\Omega_3-\omega_4)\,\Sigma_K^{a}(\epsilon_2-\Omega_3-\omega_4)
+ G^{(0)r}(\epsilon_2-\Omega_3-\omega_4)\, 
\nonumber \\ & \qquad\qquad\qquad\qquad\times
i\left(
\Sigma_K^{>}(\epsilon_2-\Omega_3-\omega_4) -
\Sigma_K^{<}(\epsilon_2-\Omega_3-\omega_4) \right)
\big]
\nonumber \\ & + 
A^{(0)}(\epsilon_2-\Omega_3-\omega_4)\,\mu\, A^{(0)}(\epsilon_2)\,\mu\left[
G^{(0)<}(\epsilon_1+\Omega_1+\omega_1)\,\mu^\dagger\, G^{(0)>}(\epsilon_1)
- (>\,\leftrightarrow\, <)
\right]\mu^\dagger
\nonumber \\ & \qquad\qquad\qquad\times
\big[
A^{(0)}(\epsilon_2-\Omega_3-\omega_4)\,\Sigma_K^{a}(\epsilon_2-\Omega_3-\omega_4)
+ G^{(0)r}(\epsilon_2-\Omega_3-\omega_4)\, 
\nonumber \\ & \qquad\qquad\qquad\qquad\times
i\left(
\Sigma_K^{>}(\epsilon_2-\Omega_3-\omega_4) -
\Sigma_K^{<}(\epsilon_2-\Omega_3-\omega_4) \right)
\big]
\nonumber \\ & + 
A^{(0)}(\epsilon_1)\,\mu\, A^{(0)}(\epsilon_1+\Omega_3+\omega_4)\,\mu^\dagger\left[
G^{(0)<}(\epsilon_2)\,\mu^\dagger\, G^{(0)>}(\epsilon_2-\Omega_1-\omega_1)
- (>\,\leftrightarrow\, <)
\right] \mu 
\nonumber \\ & \qquad\qquad\qquad\times
\big[
A^{(0)}(\epsilon_1)\,\Sigma_K^{a}(\epsilon_1) + G^{(0)r}(\epsilon_1)\,
\left(\Sigma_K^{>}(\epsilon_1)-\Sigma_K^{<}(\epsilon_1)\right)
\big]
\nonumber \\ & - 
A^{(0)}(\epsilon_1)\,\mu\, A^{(0)}(\epsilon_1-\Omega_3-\omega_4)\,\mu^\dagger
A^{(0)}(\epsilon_2)\,\mu
\nonumber \\ & \qquad\qquad\qquad\times
\big[
G^{(0)<}(\epsilon_1+\Omega_1+\omega_1)\,\mu^\dagger\left(
G^{(0)>}(\epsilon_1)\,\Sigma_K^{a}(\epsilon_1) + 
G^{(0)r}(\epsilon_1)\Sigma_K^{>}(\epsilon_1) \right)
- (>\,\leftrightarrow\, <) \big]
\bigg\}
\nonumber
\end{align}
for two-dimensional electron spectroscopy, and
\begin{align}
\label{SpOm}
 S_p(\Omega_1,T_2,\Omega_3) &= -2\,\mbox{Im}
\int\frac{d\epsilon_1}{2\pi}\int\frac{d\epsilon_2}{2\pi}\, 
E_4^{*}\, E_3\, E_2^{*}\, E_1\,
e^{iT_2(\omega_1-\omega_2+E_1-E_2)+i\Delta\phi}\,
\mbox{Tr}\bigg\{
\\ & \,\,\,\,\,\,
A^{(0)\, >}(\epsilon_2-\Omega_1-\omega_1)\,\mu
\left[G^{(0) <}(\epsilon_1)\,\mu^\dagger\, G^{(0)>}(\epsilon_2-\Omega_3-\omega_4)
- (<\,\leftrightarrow\, >)
\right]
\mu\, A^{(0)}(\epsilon_2)\,\mu^\dagger
\nonumber \\ & -
A^{(0)}(\epsilon_1)\,\mu\, A^{(0)}(\epsilon_2-\Omega_3-\omega_4)\,\mu
\left[
G^{(0)<}(\epsilon_2)\,\mu^\dagger\, G^{(0)>}(\epsilon_1+\Omega_1+\omega_1)
- (<\,\leftrightarrow\, >)
\right]\mu^\dagger
\nonumber \\ & -
\left[
G^{(0)<}(\epsilon_2-\Omega_1-\omega_1)\,\mu^\dagger\, G^{(0)>}(\epsilon_1)
- (<\,\leftrightarrow\, >)
\right]\mu\, A^{(0)}(\epsilon_1+\Omega_3+\omega_4)\,\mu\,
A^{(0)}(\epsilon_2)\,\mu^\dagger
\nonumber \\ & +
A^{(0)}(\epsilon_1)\,\mu\,\left[
G^{(0)<}(\epsilon_1+\Omega_3+\omega_4)\,\mu^\dagger\, G^{(0)>}(\epsilon_2)
- (<\,\leftrightarrow\, >)
\right]\mu\, A^{(0)}(\epsilon_1+\Omega_1+\omega_1)\,\mu^\dagger
\bigg\}
\nonumber
\end{align}
for two-dimensional photon spectroscopy.
\end{widetext}
Here, $>$, $<$, $r$, and $a$ are greater, lesser, retarded, and advanced projections,
respectively, $\Delta\phi\equiv\phi_1-\phi_2+\phi_3-\phi_4$,
\begin{equation}
A^{(0)}_{mm'}(\epsilon)\equiv i\left[G^{(0)>}_{mm'}(\epsilon)-G^{(0)<}_{mm'}(\epsilon)\right]
\end{equation}
is the spectral function.
In (\ref{SKOm})-(\ref{SpOm}) restriction 
\begin{equation}
 \omega_1-\omega_2+\omega_3-\omega_4=0
\end{equation}
is assumed to be fulfilled.
Expressions (\ref{SKOm}) and (\ref{SpOm}) are central results
of our consideration.


\section{Numerical results}\label{numres}
\begin{figure}[t]
\centering\includegraphics[width=\linewidth]{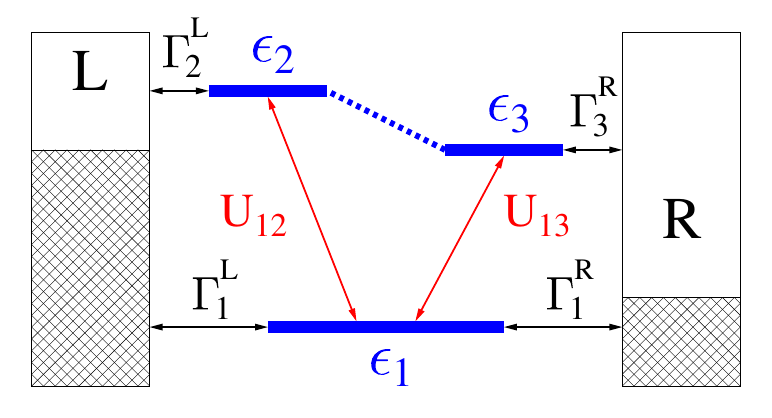}
\caption{\label{fig_model}
 (Color online) Generic junction model.
}
\end{figure}

We now illustrate two-dimensional spectroscopy signals (\ref{SKOm}) and (\ref{SpOm})
for a generic junction model. We consider molecule with two energetically different
optical transitions between ground and two excited states.
Ground state of the molecule is strongly coupled to both contacts. 
Its excited states are coupled one to left and the other to  right contact. 
Electron transfer is possible between the two excited 
states (see Fig.~\ref{fig_model})
\begin{equation}
\begin{split}
\hat H_M^{(0)} &= \sum_{m=1}^{3}\epsilon_m\hat d_m^\dagger\hat d_m
+ t_{23}\left(\hat d_2^\dagger\hat d_3+\hat d_3^\dagger\hat d_2\right)
\\
\mathbf{\mu} &= \begin{pmatrix}
0 & U_{12} & U_{13} \\
U_{12}^{*} & 0 & 0 \\
U_{13}^{*} & 0 & 0 
\end{pmatrix}
\end{split}
\end{equation}
For simplicity, we assume wide-band approximation for self-energies due to coupling to
contacts.

Parameters of the simulations are the following.
 Temperature is $k_BT=0.25$, positions of levels are 
 $\epsilon_1=-5$, $\epsilon_2=5$, $\epsilon_3=3$, and
 electron transfer matrix element $t=0.1$. Escape rates to contacts
 are $\Gamma^L_1=\Gamma^R_1=1$ and $\Gamma^L_2=\Gamma^R_3=0.1$.
 Light-matter coupling strength is $U_{12}E_\alpha=U_{13}E_\alpha=0.1$.
 Simulations are performed on energy grid spanning range from 
 $-3$ to $3$ with step $0.01$.
 Fermi energy is taken as origin, $E_F=0$, and bias $V$ is applied symmetrically,
 $\mu_{L,R}=E_F\pm \lvert e\rvert V/2$.
Here, energy is given in terms of an arbitrary unit of energy $E_0$.
Resulting signal (photon and electron fluxes) is presented in terms of flux 
unit $I_0 = 1/t_0$, where $t_0 = 1/E_0$ is unit of time.
  
\begin{figure*}[t]
\centering\includegraphics[width=\linewidth]{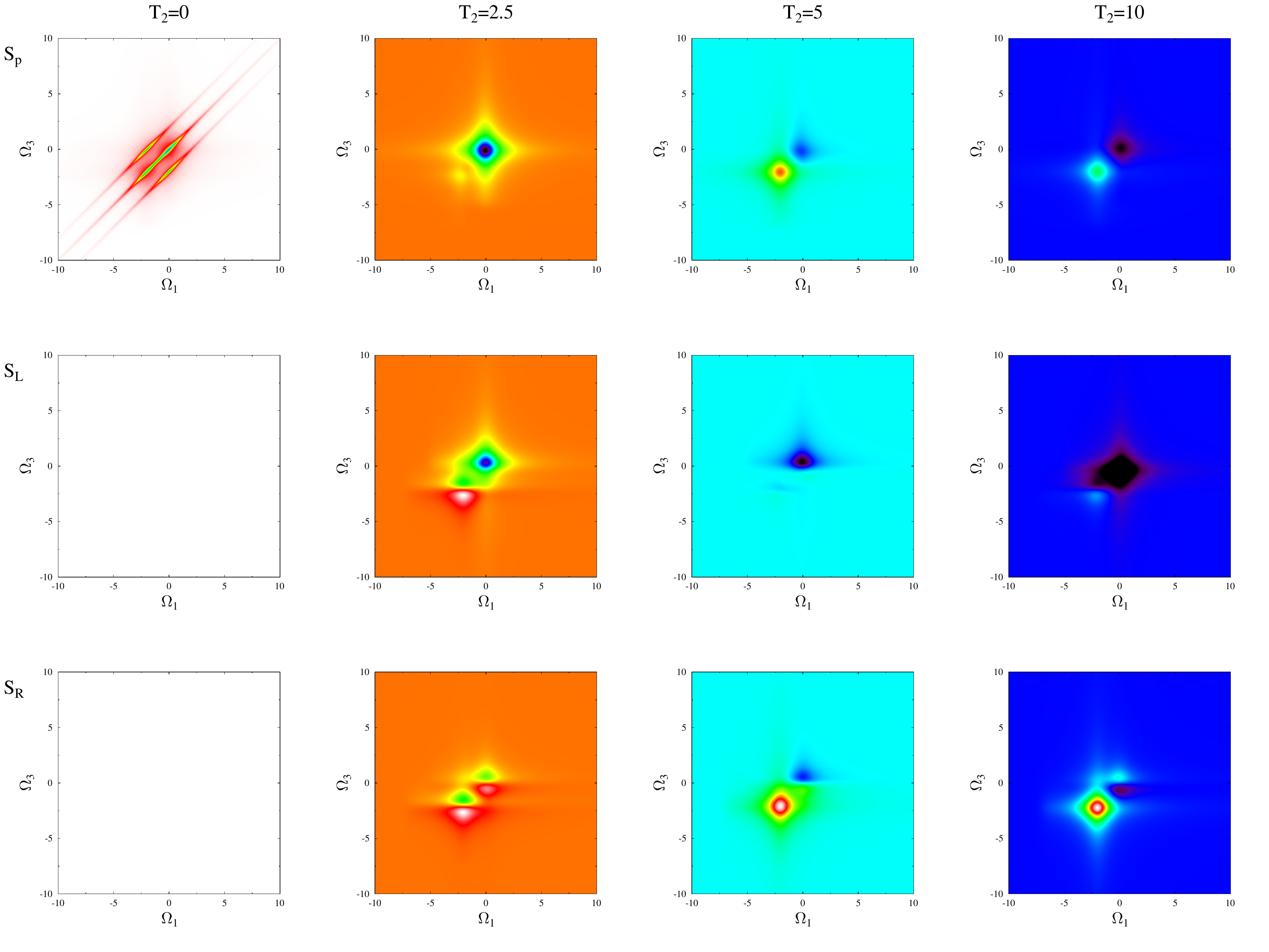}
\caption{\label{fig_S0V}
 (Color online) Two-dimensional spectroscopy for mode of Fig.~\ref{fig_model}
 at zero bias. Shown are $S_p$ (top row), $S_L$ (middle row), and $S_R$
 (bottom row) for $T_2=0$ (first column), $2.5$ (second column),
 $5$ (third column), and $10$ (last column).
}
\end{figure*}

Figure~\ref{fig_S0V} shows results of simulation employing model of Fig.~\ref{fig_mol}
at equilibrium for a set of delay times $T_2$. 
One sees that the very start only optical response $S_p$ gives usual 
two-dimensional signal with two different transitions (peaks on diagonal)
and coherences between them (off-diagonal peaks). Curren spectroscopy
does not provide any information due to the fact that at $T_2=0$
optical excitation has not affected current yet. Thus, $I_L=I_R=0$
which corresponds to zero two-dimensional signal.
As time delay increases two processes take place:
dephasing due to coupling to contacts and Rabi oscillation
affecting populations of the excited states $\epsilon_2$ and $\epsilon_3$.
Also, the excited states populations create electron current
into the left and right contacts. Thus, in optical signal 
coherences between the two excitation disappear
and relative intensity of the two peaks is defined by
Pauli blockade of one of the transitions due to Rabi oscillation 
of the excited states populations. Two-dimensional current
signal appears at finite $T_2$ with $S_L$ and $S_R$
mostly giving local information (that is, information about transition
closest to the corresponding contact). 
Note that increasing $T_2$ diminishes overall signal strength
(not shown) due to system coming closer to steady state after
disturbance by first pair of pulses.

\begin{figure*}[t]
\centering\includegraphics[width=\linewidth]{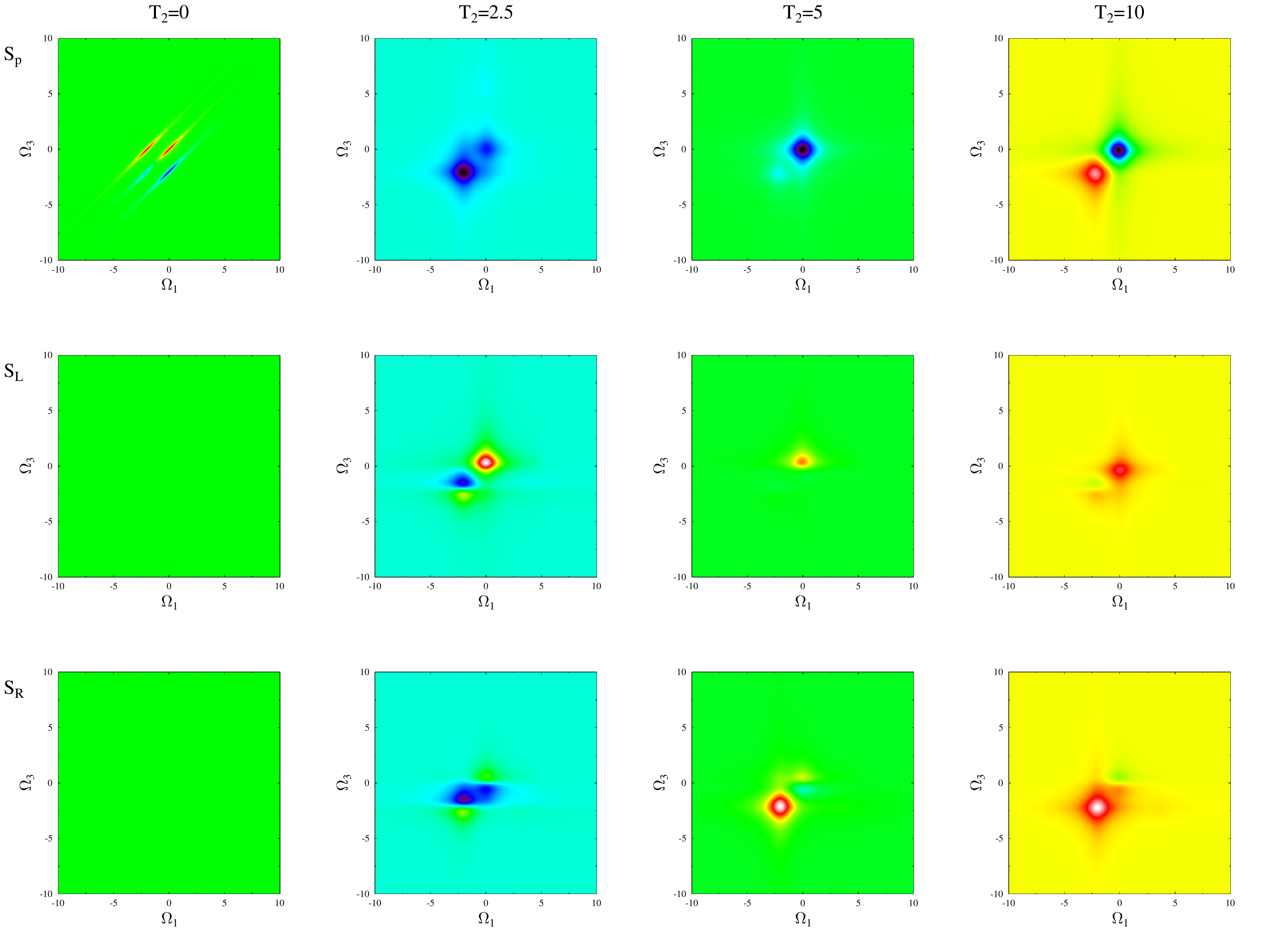}
\caption{\label{fig_S2V}
 (Color online) Two-dimensional spectroscopy for mode of Fig.~\ref{fig_model}
 at bias $\lvert e\rvert V=20$. Shown are $S_p$ (top row), $S_L$ (middle row), and $S_R$
 (bottom row) for $T_2=0$ (first column), $2.5$ (second column),
 $5$ (third column), and $10$ (last column).
}
\end{figure*}

Figure~\ref{fig_S2V} shows results of simulation employing model of Fig.~\ref{fig_model}
for biased junction, $\lvert e\rvert V=20$, for a set of delay times $T_2$. 
Qualitative behavior is the same as in equilibrium case with 
bias induced depletion of the ground state and population of excited
states affecting strength of the signal.

\begin{figure}[htbp]
\centering\includegraphics[width=\linewidth]{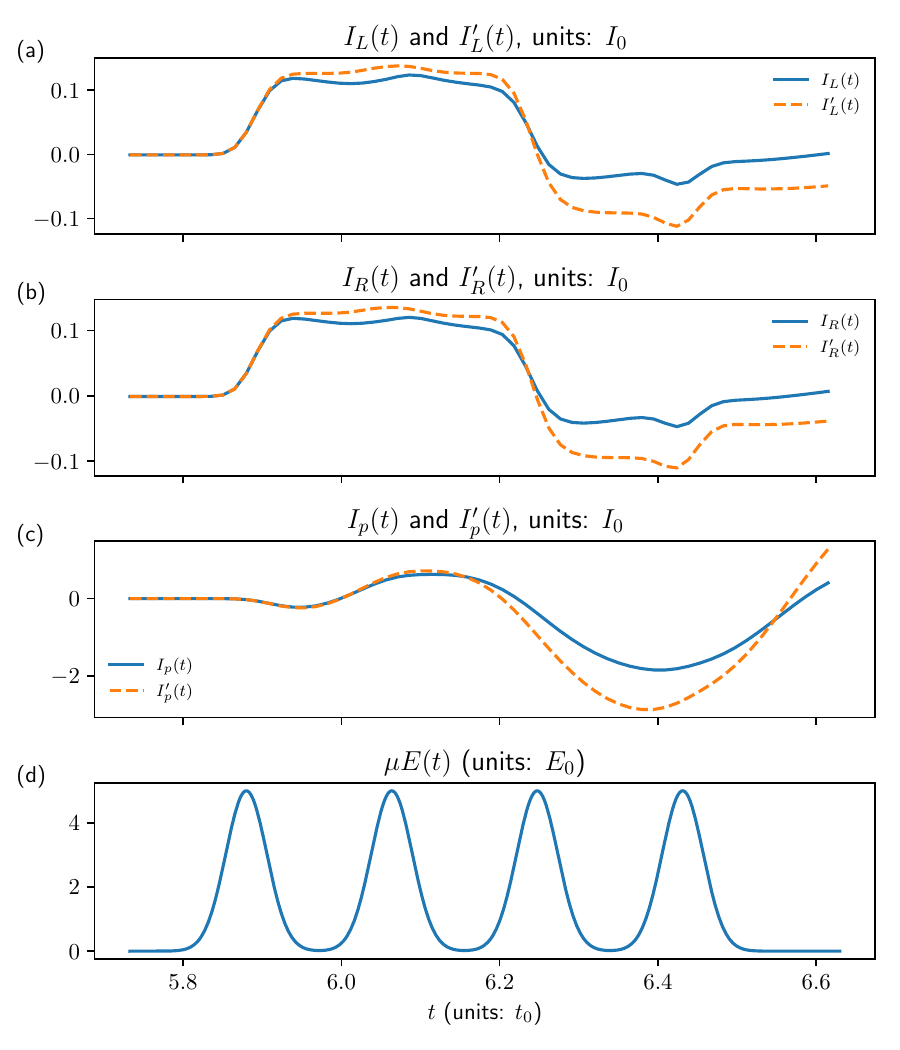}
\caption{\label{fig_currents}
 (Color online) Electron and photon fluxes induced by sequence of laser 
 pulses for junction model of Fig.~\ref{fig_model} at zero bias.
 Shown are results of full (solid line, blue) and fourth order in light-mater interaction
 (dashed line, red) NEGF simulations for (a) electron current at interface $L$,
 (b) electron current at interface $R$, and (c) photon flux.
 Panel (d) shows sequence of driving laser pulses.
 }
\end{figure}

Finally, we note that for realistic parameters traditional perturbation
theory treatment of two-dimensional spectroscopy may sometimes be 
inaccurate because it misses effect of radiation field on electronic
structure of the system. 
Figure~\ref{fig_currents} illustrates this point by comparing results of full NEGF simulation
(solid line, blue) to fourth order in light-matter interaction treatment.
Coupling to the field is $\mu E_\alpha=5$, other parameters are the same
as in previous simulations. One sees that full calculation yields results
which deviate significantly from the fourth order perturbation theory treatment.
In such cases generalization of traditional treatment
for multidimensional optical spectroscopy presented above is not enough,
and full fluxes simulation is required.

\section{Conclusions}\label{conclude}
Development of experimental techniques at nanoscale made possible
simultaneous measurements of optical (photon flux) and transport
(electron current) characteristics of open nonequilibrium quantum systems.
Theoretical description of such experiments requires generalization of
standard tools of nonlinear optical spectroscopy. In particular, formulations
of experimentally measurable bias-induced electroluminescence and Raman
spectroscopy in single-molecule junctions were discussed in the 
literature~\cite{galperin_raman_2009,galperin_raman_2011,white_raman_2014,miwa_many-body_2019}.

Two-dimensional spectroscopy is a techniques that allows to study system
responses and coherences between its degrees of freedom 
in complex molecular systems in condensed phase. Optical
spectroscopy uses photon flux to discuss system 
responses~\cite{agarwalla_coherent_2015,gao_simulation_2016}. 
Recently, experimental measurements of electron current
as the signal for the two-dimensional spectroscopy analysis were
reported in the literature~\cite{nardin_multidimensional_2013,karki_coherent_2014}.
Natural next step is two-dimensional spectroscopy simultaneously
in photon and electron fluxes.

Because photon flux and electron current are interdependent
in an open quantum system, theoretical consideration of simultaneous
two-dimensional spectroscopy requires  treatment of both fluxes
on equal footing.
Here we present theoretical formulation of two-dimensional spectroscopy 
in open quantum systems with simultaneous measurement of photon flux 
and electron current. The formulation is based on recently introduced
by us~\cite{mukamel_nonlinear_2024} 
non-self-consistent nonequilibrium Green's function approach
to nonlinear spectroscopy. Starting from this quantum mechanical
formulation we consider transition to classical description of radiation field
and use the methodology in derivation of the two-dimensional 
spectroscopy signals. The approach accounts for all the fluxes in
the system in a consistent way which (in particular) preserves conservation laws.
Expressions for the two-dimensional signals are derived as
fourth order in light-matter interaction contributions to the fluxes.

We illustrate the theoretical formulation with numerical simulations for 
a generic model of molecular junction.
We find that traditional optical signal of two-dimensional spectroscopy, $S_p$,
and electron fluxes, $S_L$ and $S_R$, provide complementary
information on the system when detected simultaneously.
In particular, for the model of FIg.~\ref{fig_model}
$S_p$ yields overall information on optical transitions,
while $S_L$ and $S_R$ represent corresponding local contributions.
Also, when excited by laser field, optical signal is available 
at earlier times than that caused by electron flux.  
Finally, we note that for realistic parameters traditional perturbation
theory treatment of two-dimensional spectroscopy may sometimes be 
inaccurate because it misses effect of radiation field on electronic
structure of the system. In such cases full fluxes simulation is required.

\begin{acknowledgments}
This material is based upon work supported by the National Science Foundation
  under Grant No. CHE-2154323.
\end{acknowledgments}


\appendix
\begin{widetext}
\section{Fourth order expressions for fluxes}\label{app_flux}
Fourth order expression for electron flux is
\begin{align}
\label{IK4}
I^{K\,(4)}_e(t) &= 
2\,\mbox{Re}\int_c d\tau' \int_c d\tau_a \int_c d\tau_b \int_c d\tau_c \int_c d\tau_d\,
\mathcal{E}^{*}(t_d)\,\mathcal{E}(t_c)\,\mathcal{E}^{*}(t_b)\,\mathcal{E}(t_a)
\\ &
\mbox{Tr}\big[
G^{(0)}(t,\tau_a)\,\mu^\dagger\, G^{(0)}(\tau_a,\tau_c)\,\mu^\dagger\, G^{(0)}(\tau_c,\tau_b)\,
\mu\,\,\,\, G^{(0)}(\tau_b,\tau_d)\,\mu\,\,\,\, G^{(0)}(\tau_d,\tau')\, \Sigma^K(\tau',t)
\nonumber \\ & \,\,\,
+ G^{(0)}(t,\tau_a)\,\mu^\dagger\, G^{(0)}(\tau_a,\tau_d)\,\mu\,\,\,\, G^{(0)}(\tau_d,\tau_b)\,
\mu\,\,\,\, G^{(0)}(\tau_b,\tau_c)\,\mu^\dagger\, G^{(0)}(\tau_c,\tau')\, \Sigma^K(\tau',t)
\nonumber \\ & \,\,\,
+ G^{(0)}(t,\tau_b)\,\mu\,\,\,\, G^{(0)}(\tau_b,\tau_c)\,\mu^\dagger \, G^{(0)}(\tau_c,\tau_a)\,
\mu^\dagger\, G^{(0)}(\tau_a,\tau_d)\,\mu\,\,\,\, G^{(0)}(\tau_d,\tau')\, \Sigma^K(\tau',t)
\nonumber \\ & \,\,\,
+ G^{(0)}(t,\tau_b)\,\mu\,\,\,\, G^{(0)}(\tau_b,\tau_d)\,\mu\,\,\,\, G^{(0)}(\tau_d,\tau_a)\,
\mu^\dagger\, G^{(0)}(\tau_a,\tau_c)\,\mu^\dagger\, G^{(0)}(\tau_c,\tau')\, \Sigma^K(\tau',t)
\nonumber \\ & \,\,\,
+ G^{(0)}(t,\tau_a)\,\mu^\dagger\, G^{(0)}(\tau_a,\tau_b)\,\mu\,\,\,\, G^{(0)}(\tau_b,\tau_c)\,
\mu^\dagger\, G^{(0)}(\tau_c,\tau_d)\,\mu\,\,\,\, G^{(0)}(\tau_d,\tau') \,\Sigma^K(\tau',t)
\nonumber \\ & \,\,\,
+ G^{(0)}(t,\tau_a)\,\mu^\dagger\, G^{(0)}(\tau_a,\tau_b)\,\mu\,\,\,\, G^{(0)}(\tau_b,\tau_d)\,
\mu\,\,\,\, G^{(0)}(\tau_d,\tau_c)\,\mu^\dagger\, G^{(0)}(\tau_c,\tau') \,\Sigma^K(\tau',t)
\nonumber \\ & \,\,\,
+ G^{(0)}(t,\tau_b)\,\mu\,\,\,\, G^{(0)}(\tau_b,\tau_a)\,\mu^\dagger\, G^{(0)}(\tau_a,\tau_c)\,
\mu^\dagger\, G^{(0)}(\tau_c,\tau_d)\,\mu\,\,\,\, G^{(0)}(\tau_d,\tau')\, \Sigma^K(\tau',t)
\nonumber \\ & \,\,\,
+ G^{(0)}(t,\tau_b)\,\mu\,\,\,\, G^{(0)}(\tau_b,\tau_a)\,\mu^\dagger\, G^{(0)}(\tau_a,\tau_d)\,
\mu\,\,\,\, G^{(0)}(\tau_d,\tau_c)\,\mu^\dagger\, G^{(0)}(\tau_c,\tau')\, \Sigma^K(\tau',t)
\nonumber \\ & \,\,\,
+ G^{(0)}(t,\tau_a)\,\mu^\dagger\, G^{(0)}(\tau_a,\tau_c)\,\mu^\dagger\, G^{(0)}(\tau_c,\tau_d)\,
\mu\,\,\,\, G^{(0)}(\tau_d,\tau_b)\,\mu\,\,\,\, G^{(0)}(\tau_b,\tau')\, \Sigma^K(\tau',t)
\nonumber \\ & \,\,\,
+ G^{(0)}(t,\tau_a)\,\mu^\dagger\, G^{(0)}(\tau_a,\tau_d)\,\mu\,\,\,\, G^{(0)}(\tau_d,\tau_c)\,
\mu^\dagger\, G^{(0)}(\tau_c,\tau_b)\,\mu\,\,\,\, G^{(0)}(\tau_b,\tau')\, \Sigma^K(\tau',t)
\nonumber \\ & \,\,\,
+ G^{(0)}(t,\tau_b)\,\mu\,\,\,\, G^{(0)}(\tau_b,\tau_c)\,\mu^\dagger\, G^{(0)}(\tau_c,\tau_d)\,
\mu\,\,\,\, G^{(0)}(\tau_d,\tau_a)\,\mu^\dagger\, G^{(0)}(\tau_a,\tau')\, \Sigma^K(\tau',t)
\nonumber \\ & \,\,\,
+ G^{(0)}(t,\tau_b)\,\mu\,\,\,\, G^{(0)}(\tau_b,\tau_d)\,\mu\,\,\,\, G^{(0)}(\tau_d,\tau_c)\,
\mu^\dagger\, G^{(0)}(\tau_c,\tau_a)\,\mu^\dagger\, G^{(0)}(\tau_a,\tau')\, \Sigma^K(\tau',t)
\big]
\nonumber
\end{align}
Fourth order expression for photon flux is
\begin{align}
\label{Ip4}
I_p^{(4)}(t) &= 2\,\mbox{Re} \int_c d\tau_a \int_c d\tau_b \int_c d\tau_c\,
\mathcal{E}^{*}(t)\,\mathcal{E}(t_c)\,\mathcal{E}^{*}(t_b)\,\mathcal{E}(t_a)
\\ & \mbox{Tr}\big[
G^{(0)}(t,\tau_a)\,\mu^\dagger\, G^{(0)}(\tau_a,\tau_b)\, \mu\,\,\,\, G^{(0)}(\tau_b,\tau_c)\,
\mu^\dagger\, G^{(0)}(\tau_c,t)\,\mu
\nonumber \\ & \,\,\,
+ G^{(0)}(t,\tau_b)\,\mu\,\,\,\, G^{(0)}(\tau_b,\tau_a)\, \mu^\dagger\, G^{(0)}(\tau_a,\tau_c)\,
\mu^\dagger\, G^{(0)}(\tau_c,t)\,\mu
\nonumber \\ & \,\,\,
+ G^{(0)}(t,\tau_c)\,\mu^\dagger\, G^{(0)}(\tau_c,\tau_a)\, \mu^\dagger\, G^{(0)}(\tau_a,\tau_b)\,
\mu\,\,\,\, G^{(0)}(\tau_b,t)\,\mu
\nonumber \\ & \,\,\,
+ G^{(0)}(t,\tau_c)\,\mu^\dagger\, G^{(0)}(\tau_c,\tau_b)\, \mu\,\,\,\, G^{(0)}(\tau_b,\tau_a)\,
\mu^\dagger\, G^{(0)}(\tau_a,t)\,\mu
\nonumber \\ & \,\,\,
+ G^{(0)}(t,\tau_b)\,\mu\,\,\,\, G^{(0)}(\tau_b,\tau_c)\, \mu^\dagger\, G^{(0)}(\tau_c,\tau_a)\,
\mu^\dagger\, G^{(0)}(\tau_a,t)\,\mu
\nonumber \\ & \,\,\,
+ G^{(0)}(t,\tau_a)\,\mu^\dagger\, G^{(0)}(\tau_a,\tau_c)\, \mu^\dagger\, G^{(0)}(\tau_c,\tau_b)\,
\mu\,\,\,\, G^{(0)}(\tau_b,t)\,\mu
\big]
\nonumber 
\end{align}
\end{widetext}


%

\end{document}